\documentclass[english,comsoc,10pt]{IEEEtran}
\usepackage[T1]{fontenc}
\usepackage[latin9]{inputenc}
\usepackage{array}
\usepackage{multirow}
\usepackage{amsthm}
\usepackage{graphicx}

\makeatletter

\providecommand{\tabularnewline}{\\}

  \theoremstyle{plain}
  \newtheorem{thm}{\protect\theoremname}
  \theoremstyle{remark}
  \newtheorem{rem}{\protect\remarkname}

\usepackage{geometry}
\usepackage{mathptm}
\makeatother

\usepackage{babel}
\providecommand{\remarkname}{Remark}
\providecommand{\theoremname}{Theorem}

\begin{document}

\title{Quantitative Models of Imperfect Deception in Network Security using
Signaling Games with Evidence {[}IEEE CNS 17 Poster{]}}

\author{Jeffrey Pawlick and Quanyan Zhu }
\maketitle
\begin{abstract}
Deception plays a critical role in many interactions in communication
and network security. Game-theoretic models called ``cheap talk signaling
games'' capture the dynamic and information asymmetric nature of
deceptive interactions. But signaling games inherently model undetectable
deception. In this paper, we investigate a model of signaling games
in which the receiver can detect deception with some probability.
This model nests traditional signaling games and complete information
Stackelberg games as special cases. We present the pure strategy perfect
Bayesian Nash equilibria of the game. Then we illustrate these analytical
results with an application to active network defense. The presence
of evidence forces majority-truthful behavior and eliminates some
pure strategy equilibria. It always benefits the deceived player,
but surprisingly sometimes also benefits the deceiving player. 
\end{abstract}

\section{Introduction\label{sec:intro}}

\begin{figure}
\begin{centering}
\includegraphics[width=1\columnwidth]{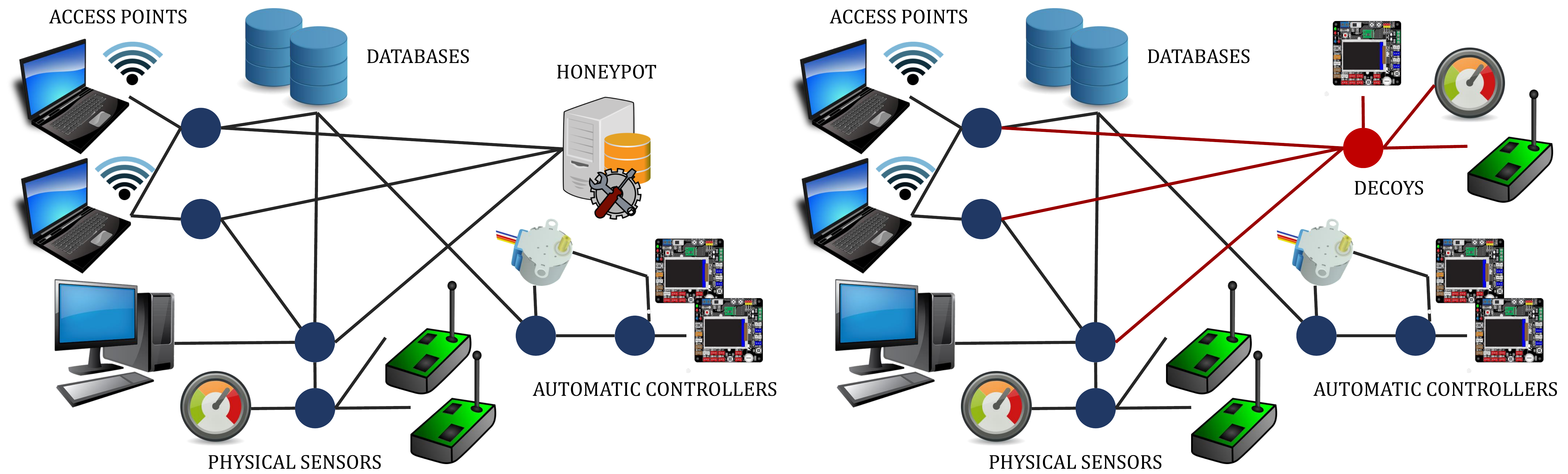}
\par\end{centering}

\caption{\label{fig:activeCybDef}An example of defensive deception. The left
side shows the real network and the right side shows the deceptive
network. A honeynet is disguised as a set of sensors and a controller
in order to manipulate attacker movements. This interaction can be
modeled by signaling games with evidence.}
\end{figure}

Advanced cyberattackers employ deception to evade signature detection,
release misleading information, and frustrate attempts at attribution.
Deceptive opinion spam \cite{ott2011finding} and identity deception
in social networks \cite{tsikerdekis2017identity} are two examples.
Deception can also be used in \emph{active cyber defense} to manipulate
the beliefs of an adversary \cite{stech2016integrating}, leveraging
the advantage of information-asymmetry typically enjoyed by attackers
(Fig. \ref{fig:activeCybDef}).

Quantitative metrics are needed to optimally deploy defensive deception
and optimally detect and mitigate malicious deception. These metrics
would also allow policymakers, entrepreneurs, and cyber-insurance
vendors to assess the influence of new legislation, technology, or
risk mitigation strategies. Game theory provides a set of tools to
make quantitative, verifiable predictions about the outcome of the
strategic and decentralized decisions characteristic of network security.
In particular, \emph{cheap talk signaling games} \cite{crawford1982strategic}
capture the dynamic and information-asymmetric nature of deceptive
interactions. These games are two-player, dynamic, information asymmetric
games. The players are a sender ($S$) and a receiver ($R$), which
correspond to the party which may attempt deception and the party
which may be deceived, respectively. 

Cheap talk signaling games are often used to model deception in cybersecurity.
But these games inherently model deception which is \emph{undetectable}\footnote{In some signaling games, equilibrium conditions allow the message
to convey the true private information. But there is no exogenous
constraint on deception, \emph{i.e.}, it is just as easy for the sender
to lie as it is for him to reveal the truth. }\emph{.} Of course, both security administrators and cybercriminals
invest heavily in detecting deception. Examples include detection
of false opinion spam \cite{ott2011finding}, malicious logins \cite{siadati2016detecting},
and social network identity deception \cite{tsikerdekis2017identity}.
Therefore, we extend cheap-talk signaling games to capture the possibility
of detecting deception. 

\section{Model\label{sec:model}}

\begin{figure}
\begin{centering}
\includegraphics[width=0.75\columnwidth]{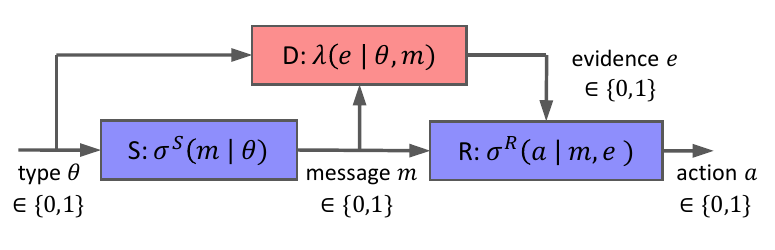}
\par\end{centering}

\caption{\label{fig:sigWEv}In signaling games with evidence, $R$ chooses
action $a$ with a probability that depends on evidence $e$ emitted
by the detector $D$ in addition to message $m.$ }
\end{figure}
Figure \ref{fig:sigWEv} depicts the traditional signaling game between
$S$ and $R,$ augmented by a detector block ($D$). We can call this
augmented signaling game a \emph{signaling game with evidence} \cite{pawlick2015deception}.
Our contribution is to add block $D,$ which denotes a \emph{detector}
that emits evidence $e\in E=\left\{ 0,1\right\} $ with probability
$\lambda\left(e\,|\,\theta,m\right).$ The detector classifies the
message as \emph{suspicious }($e=1$) or \emph{not suspicious} ($e=0$).
Two examples of detectors are email clients which warn users about
possible phishing emails and browser warnings which alert users if
websites do not have verifiable website security certificates. Let
$\beta\in[1/2,1]$ and $\alpha\in[0,1/2]$ denote the \emph{power}
and \emph{size} of the detector, respectively. $R$ uses both the
message $m$ and the evidence $e$ to form belief $\mu(\theta\,|\,m,e)$
about the likelihood that $S$ has type $\theta.$

\section{Analytical Results and Application}

Theorem \ref{thm:pbne} states the perfect Bayesian Nash equilibria
(PBNE) \cite{fudenberg1991game} of the game. Remark \ref{rem:forceTruth}
and Remark \ref{rem:stackMid} discuss important properties of the
PBNE, especially focusing on the ways in which signaling games with
evidence differ from traditional signaling games.
\begin{thm}
\label{thm:pbne}Table \ref{tab:PBNEsummary} summarizes the pure
strategy PBNE \footnote{Algebraic quantities for strategies and beliefs on and off the equilibrium
path, as well as results when $\beta<1-\alpha,$ have also been obtained
but are not presented here due to space limitations.}. \end{thm}
\begin{rem}
\label{rem:forceTruth} Without evidence, it is equivalent for $S$
to always reveal the truth or to always lie. One effect of evidence
is to force \emph{majority-truthful} signaling. In the \texttt{0-Majority}
regime, $S$ of type $0$ reveal truthfully and $S$ of (the minority)
type $1$ deceive. The opposite occurs in the \texttt{1-Majority}
regime.
\end{rem}
\begin{table}
\caption{\label{tab:PBNEsummary}Pure Pooling PBNE when $\beta>1-\alpha$}

\centering{}%
\begin{tabular}{|c|c|c|}
\hline 
Prior Probabilities & Sender w/o Evidence & Sender w/ Evidence\tabularnewline
\hline 
\multirow{1}{*}{\texttt{0-Dominant}} & Reveal or deceive & Reveal or deceive\tabularnewline
\hline 
\multirow{1}{*}{\texttt{0-Majority}} & Reveal or deceive & \multirow{1}{*}{Majority reveal}\tabularnewline
\hline 
\multirow{1}{*}{\texttt{Mixed}} & Reveal or deceive & \multirow{1}{*}{No Eq.}\tabularnewline
\hline 
\multirow{1}{*}{\texttt{1-Majority}} & Reveal or deceive & \multirow{1}{*}{Majority reveal}\tabularnewline
\hline 
\multirow{1}{*}{\texttt{1-Dominant}} & Reveal or deceive & Reveal or deceive\tabularnewline
\hline 
\end{tabular}
\end{table}

\begin{rem}
\label{rem:stackMid}In the \texttt{Mixed} prior probability regime,
the evidence eliminates all PBNE by playing a dominant role. Here
$R$ trusts $S$ if $e=0$ and does not trust $S$ if $e=1.$ It can
be shown that $S$ and $R$ can never mutually counter each other's
strategies.
\end{rem}
Now consider an application in which a network administrator $S$
is defending a network from an attacker $R$ by camouflaging normal
systems as honeypots or honeypots as normal systems. Let $\theta=0$
and $\theta=1$ denote normal systems and honeypots, respectively.
Let $m=0$ and $m=1$ denote camouflaging (or revealing) a system
as a normal system or a honeypot. But this camouflage is not perfect,
because the attacker can try to detect honeypots through tests such
as measuring the execution time of control-modifying CPU instructions
\cite{franklin2008remote}. This produces evidence $e=1$ for a suspicious
system (\emph{i.e.}, one in which it is likely that $m\neq\theta$),
and $e=0$ for a system which is not suspicious. $R$ uses this to
decide whether to move into the system or to withdraw.

\begin{figure}
\begin{centering}
\includegraphics[width=0.7\columnwidth]{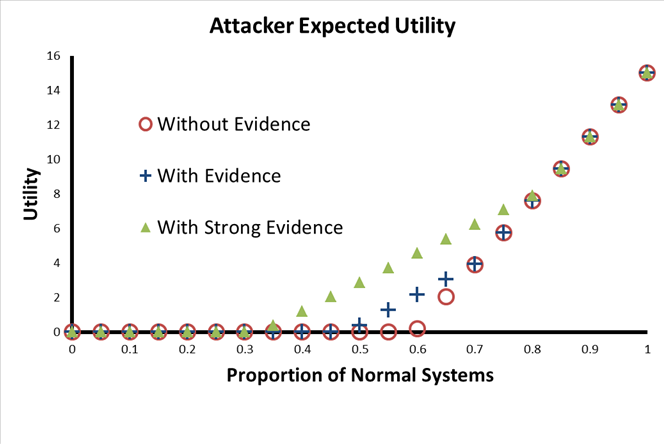}
\par\end{centering}

\caption{\label{fig:uSAttacker1}Expected utility for the attacker as a function
of the fraction $p\left(0\right)$ of normal systems in the network.}

\begin{centering}
\includegraphics[width=0.7\columnwidth]{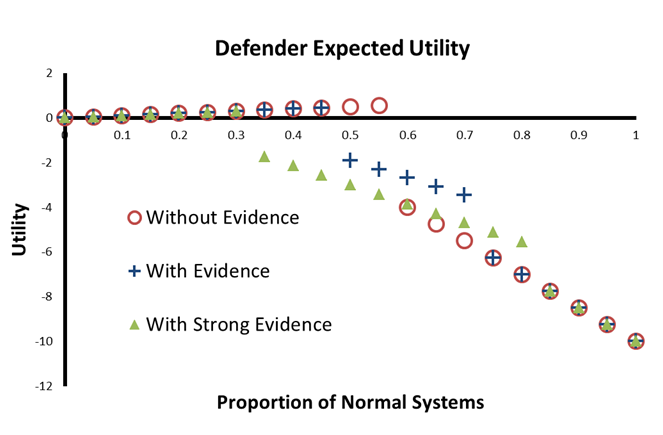}
\par\end{centering}

\caption{\label{fig:uDefender}Expected utility for the defender as a function
of the fraction $p\left(0\right)$ of normal systems in the network.}
\end{figure}

Figures \ref{fig:uSAttacker1}-\ref{fig:uDefender} use Gambit \cite{gambit2014software}
to illustrate the results. Evidence always improves the expected utility
of the attacker $R.$ He always benefits from being able to detect
honeypots. Interestingly, the defender $S$ also sometimes benefits
from evidence, as illustrated by Fig. \ref{fig:uDefender}. This implies
that she sometimes wants to imperfectly obscure the network characterization.

\section{Conclusion}

Traditional signaling games model deception which is impossible to
detect. We have introduced \emph{signaling games with evidence}, which
allow an exogenous probability of detecting deception. Evidence forces
\emph{majority-truthful }behavior in some parameter regimes. It also
eliminates all pure strategy equilibria in others. The capability
to collect evidence is always beneficial for the uniformed player.
Surprisingly, detection is sometimes advantageous to the deceiver.
We have illustrated an application to network defense using honeypots,
but our model applies to any active cybersecurity defense which imperfectly
leverages information asymmetry.

\bibliographystyle{plain}
\bibliography{SigEvBib}

\end{document}